\documentclass[epjST]{svjour}
\usepackage{amsmath,amssymb}
\usepackage{graphicx}
\usepackage{bm}
\usepackage{oldgerm}
\newcommand{\be}{\begin{equation}}
\newcommand{\ee}{\end{equation}}
\newcommand{\bee}{\begin{eqnarray}}
\newcommand{\eee}{\end{eqnarray}}

\begin{document}
\title{\boldmath 
Spin-dependent observables
and the $D_2$ parameter in breakup of deuteron and $^3$He}
\author{A.\,P.~Kobushkin\inst{1}\fnmsep\thanks{\email{kobushkin@bitp.kiev.ua}}
\and E.A.~Strokovsky\inst{2}\fnmsep\thanks{\email{strok@sunse.jinr.ru} }
\institute{Bogolyubov Institute for Theoretical Physics, Metrologicheskaya
Street 14B,
03680 Kiev, Ukraine;
and
National Technical University of Ukraine ``KPI'',
Prospekt Peremogy 37, 03056 Kiev, Ukraine.
\and
Joint Institute for Nuclear Research, Laboratory of High Energy Physics,
141980, Dubna, Russia.}}
\abstract{
We analyze the momentum
distributions of constituents in $^3$He, as well as the spin-dependent
observables for $(^3\mathrm{He},d)$, $(^3\mathrm{He},p)$, and  $(\mathrm{d},p)$
breakup reactions. Special attention is paid to the region of small relative momenta
of the helium-3 and the deuteron constituents, where a single parameter, $D_2$,
has determining role for the spin-dependent observables. We extract also this
parameter for the deuteron, basing on existing data for the tensor analyzing
power of this reaction.
}
\maketitle

\section{\label{sec:Introduction}Introduction}
Momentum distributions of one and two nucleon fragments in the lightest nuclei such
as $^3$He and deuteron give important information about nuclear system structure.
They cast light on such interesting problems as the nucleon-nucleon
interaction at short distances, the role of three-body
interaction (the 3$N$-forces in the $^3$He case), and non-nucleonic degrees of
freedom in nuclei. Data on spin-dependent observables contain an important
complementary information to this.

Precise data are currently available on the momentum distributions of the
proton and the deuteron in $^3$He obtained with
electromagnetic~\cite{em_probe1,em_probe2,em_probe3,em_probe4} and hadronic
probes~\cite{Ableev3He,Kitching,Epstein}. Data on the energy dependence of
the differential cross sections of backward elastic
$^3\text{He}(p,^{3\!}\text{He})p$ scattering, which are related to the same
momentum distributions, also exist~\cite{3hebes,RCNP}. Furthermore, the
spin-correlation parameter $C_{yy}$ for this reaction was recently measured
for first time~\cite{RCNP}. Finally, the tensor polarization of the deuteron
in the $^{12}\mathrm{C}(^3\mathrm{He},d)$ reaction was also
measured~\cite{Sitnik,Sitnikprel}. Both this and the $C_{yy}$ data~\cite{RCNP} are
sensitive to the spin structure of $^3$He.

A convenient parametrization of the fully antisymmetric three-nucleon wave
function based on the Paris~\cite{Paris} and CD-Bonn~\cite{CD-Bonn}
potentials has been presented~\cite{Baru}.  We used it in
Ref.~\cite{3hekobeas} in order to calculate the momentum
distributions in $^3$He, as well as the spin-dependent observables, within the
framework of the spectator model for the $^3$He breakup reactions. We paid
in~\cite{3hekobeas} special attention to the study of the two-body
$^3\mathrm{He}\rightarrow d+p$ channel and compared our results with other
theoretical works and existing experimental data.

In our analysis~\cite{3hekobeas} of spin-dependent observables
for $(^3\mathrm{He},d)$ and $(^3\mathrm{He},p)$ reactions, we carefully consider
their behavior in the region of small (below $\approx 150$\,MeV/$c$) internal momenta
of the $^3$He fragments, where a single quantity, known in the literature as
the $D_2$ parameter, completely determines both the sign and the momentum dependence
of the observables.

Similar parameter is known for the bound 2$N$ system (the deuteron) as well.
It determines the behavior of spin-dependent observables for the $(\mathrm{d},p)$
breakup in the same sense as for the $^3$He case, but for the $(\mathrm{d},p)$
breakup rather good database exists what makes possible an independent extraction
of this parameter. We performed here the corresponding analysis; the obtained
result agrees well with existing theoretical values as well as with experimental
estimations, extracted from low energy reactions.

\section{\label{Sec:2}The parametrization of the three-nucleon wave function}

We here give a brief review of the parametrization of the $\mathrm{^3He}$ wave
function~\cite{Baru}. Working in the framework of the so-called channel spin
coupling scheme (Ref.~\cite{Schadow}), the authors of Ref.~\cite{Baru}
restricted themselves to five partial waves
\be
\label{Channelspin}
\left|\left[\left((\ell s)j\tfrac12\right)KL\right]\tfrac12\right>,
\ee
where $\ell$, $j$ and  $s$ are the orbital, total, and spin angular momenta
for the pair (the $2^{\text{nd}}$ and $3^{\text{rd}}$ nucleons), $L$ and $K$ are
relative orbital angular momentum for the spectator (the $1^{\text{st}}$
nucleon) and the channel spin, respectively. Coulomb effects are not
included. The appropriate quantum numbers of the partial waves are collected
in Table~\ref{tab:QN}.

We use the standard definition of the Jacobi coordinates $\mathbf{r}$
(the relative coordinate between nucleons in the pair) and
\mbox{\boldmath$\rho$} (the relative coordinate between the nucleon-spectator
and the pair) with the corresponding momenta being $\mathbf{p}$ and $\mathbf{q}$.

Explicitly, the wave function of $\mathrm{^3He}$ in momentum space,
normalized to unity, reads (see also Ref.~\cite{3hekobeas}):
\begin{equation}\label{Explicity}
\begin{split}
&\Psi_\sigma(\mathbf p, \mathbf q\,)  =
\sum_\xi\left\{\frac{1}{4\pi}\delta_{\xi\sigma}
\sum_{\tau_3,t_3}
\left<1\tfrac12 \tau_3t_3\bigm|\tfrac12\tfrac12\right>\psi_1(p,q)
\left|00;1\tau_3\right>\chi_{\xi t_3} \right. \\
&\left. + \sum_{s_3}\left[
\frac{1}{4\pi}\left<1\tfrac12s_3\xi\bigm|\tfrac12\sigma\right> \psi_2(p,q) 
- \sqrt{\frac{1}{4\pi}}\sum_{L_3K_3}\left<1\tfrac12s_3\xi\bigm|\tfrac32 K_3\right>
\left<\tfrac32 2 K_3 L_3 \bigm|\tfrac12\sigma\right> \right. \right. \\
& \times Y_{2L_3}(\widehat {\mathbf q}\,)\psi_3(p,q) 
\left.\left.
-\sqrt{\frac{1}{4\pi}}\sum_{\ell_3M}\left<12 s_3\ell_3|1M\right>
\left<1\tfrac12M\xi\bigm|\tfrac12 \sigma\right>  Y_{2\ell_3}(\widehat {\mathbf p}\,)\psi_4(p,q)\right.\right. \\
&\left.\left.
+\sum_{\ell_3ML_3K_3}\left<12 s_3\ell_3|1M\right>
\left<1\tfrac12M\xi\bigm|\tfrac32 K_3\right>
\left<\tfrac32 2 K_3 L_3\bigm|\tfrac12\sigma \right> 
 Y_{2L_3}(\widehat {\mathbf q}\,) Y_{2\ell_3}(\widehat {\mathbf p}\,) 
\right. \right. \\
& \left. \phantom{\sum_{\tau_3,t_3}} \left. \phantom{\frac{1}{4\pi}}\times
\psi_5(p,q) \right] \left|1s_3;00\right>\chi_{\xi\frac12}\right\} \  {,}
\end{split}
\end{equation}
where $\sigma$ and $\xi$ are the spin projections of $^3$He and the
nucleon-spectator, $t_3$ is the isospin projection of the nucleon-spectator,
$M$ is the projection of the total angular momentum of the pair, $\chi_{\xi
t_3}$ and $\left|ss_3;\tau\tau_3\right>$ are the spin-isospin wave function
of the spectator nucleon and the pair, respectively.

The values of the partial channel probabilities, defined as
$P_\nu=\frac{1}{3}\int d^3q\,\rho_{\nu}(q)= \int
dp\,dq\,p^2q^2|\psi_\nu(p,q)|^2$,
are given in the last two columns of Table~\ref{tab:QN}.

It is important to note that the distributions for the
$^1s_0S$ and $^3s_1S$ channels are very similar in both their magnitude and
their momentum dependence.

\begin{table*}[h]
\caption{\label{tab:QN}Quantum numbers of the $^3\mathrm{He}$ partial waves. Here
$s$, $\tau$, $\ell$ and $j$ are spin, isospin, orbital and total
angular momenta of the pair; $L$ and $K$ are relative angular momenta for the
spectator and the channel spin, respectively.}
\begin{center}
\renewcommand{\arraystretch}{1.18}
\begin{tabular}{cccccccccc}
Channel  &  Label    &   $\ell$ & $s$& $j^\pi$ & $K$ & $L$ & $\tau$ & \multicolumn{2}{c}{$P_\nu$} \\ \cline{9-10}
  \# &     &   & &  & &  & & Paris & CD-Bonn \\
\hline
1           &  $^1s_0S$ &    $\  0\  $     & $\  0\  $  & $\  0^+\  $   & $\  1/2\  $ &  $\  0\  $  &  $\  1\  $ & 0.5000 & 0.5000 \\
2           &  $^3s_1S$ &    0     & 1  & $1^+$   & 1/2 &  0  &  0 & 0.4600 &  0.4658 \\
3           &  $^3s_1D$ &    0     & 1  & $1^+$   & 3/2 &  2  &  0 & 0.0282 & 0.0231 \\
4           &  $^3d_1S$ &    2     & 1  & $1^+$   & 1/2 &  0  &  0 & 0.0103 & 0.0102 \\
5           &  $^3d_1D$ &    2     & 1  & $1^+$   & 3/2 &  2  &  0 & 0.0015 & 0.0009 \\
\hline
\end{tabular}
\renewcommand{\arraystretch}{1.}
\end{center}
\vspace{-0.75cm}
\end{table*}

We use the following convention for angular momentum summation in Eq.~(\ref{Explicity}):
\be\label{convention}
j+\tfrac12\to K,\qquad K+L \to \tfrac12.
\ee

Other conventions are often used in the literature, for example:
\begin{eqnarray}
& & j+\tfrac12\to K,\qquad L+K \to \tfrac12\  {,} \label{alternative_convention} \\
& &\tfrac12+j\to K,\qquad L+K \to \tfrac12\  {.} \label{friar convention}
\end{eqnarray}

The convention of Eq.~(\ref{alternative_convention}) was used, in particular,
in Ref.~\cite{Knutson}, whereas that of Eq.~(\ref{friar convention}) was
exploited in Ref.~\cite{friar}.

Due to the properties of the Clebsch--Gordan coefficients under permutations,
some of the wave function
components have opposite signs in different conventions. For example, using
Eq.~(\ref{alternative_convention}) rather than Eq.~(\ref{convention}) would
result in $\psi_3(p,q) \to -\psi_3(p,q)$ and $\psi_5(p,q) \to -\psi_5(p,q)$.
Similarly, the use of Eq.~(\ref{friar convention}) instead of
Eq.~(\ref{convention}) would give $\psi_2(p,q) \to -\psi_2(p,q)$,
$\psi_3(p,q) \to -\psi_3(p,q)$, $\psi_4(p,q) \to -\psi_4(p,q)$, and
$\psi_5(p,q) \to -\psi_5(p,q)$, while $\psi_1(p,q)$ would not change sign.

\section{\label{sec:MD}Momentum distributions}
\subsection{\label{sec:OneNucl}One-nucleon distributions}

The momentum distribution of a nucleon $N$ with spin and isospin projections
$\xi$ and $t_3$ in $^3$He with spin projection $\sigma$ is
\be
\label{OneNuclMD}
N_{\sigma (\xi t_3)}(\mathbf q) =
3\sum_{ss_3\tau\tau_3}\int d^3p\left|
\chi^\dag_{\xi t_3}\left<ss_3\tau\tau_3\right|\Psi_\sigma (\mathbf p,\mathbf q)\right|^2\,.
\ee

In the neutron case, Eq.~\eqref{OneNuclMD} reduces to
$n_{\sigma \xi}(q)
=\tfrac23\delta_{\sigma \xi}\rho_1(q) \equiv \delta_{\sigma \xi}n(q)$;
the number of neutrons in $^3$He is $\mathcal N_n=\int d^3q\,n(q)=1$, so
the $\psi_{1}$ component must be normalized as
$\int dp\, dq\,p^2q^2\left[\psi_{1}(p,q)\right]^2=1/2$. 
Here and below we use $p_{\sigma \xi}$ and $n_{\sigma \xi}$ instead of
$N_{\sigma (\xi, \frac12)}$ and $N_{\sigma (\xi, -\frac12)}$, respectively.

The momentum distribution of the proton, given by the sum of
$p_{\frac12\frac12}(q,\theta)$ and $p_{\frac12-\frac12}(q,\theta)$
(where $p_{\frac12\frac12}(q,\theta)$ and $p_{\frac12-\frac12}(q,\theta)$
are the momentum distributions of protons with spin projection
$\frac12$ and $-\frac12$ in the
$^3$He having spin projection $\mathbf{+}\frac12$) is:
\be
\label{pMD}
p(q)=
\frac{1}{3}\rho_{1}(q)+\rho_{2}(q)+\rho_{3}(q)+\rho_{4}(q)+\rho_{5}(q)\ {.}
\ee

The number of protons in $^3$He is $\mathcal N_p=\int d^3q\, p(q) =2$ (see
Ref. \cite{3hekobeas}).

\subsection{Two-nucleon momentum distributions}

We define the two-body amplitudes $A_{dp}(M,\xi,\sigma,\mathbf q)$ as
\begin{eqnarray}
&A_{dp}(M,\xi,\sigma,\mathbf q) 
=(2\pi)^{\frac32}\sqrt 3\int d^3p\, {\psi_{d}}^\dag(M,\mathbf p)\chi^\dag_{\xi\frac12}
\Psi_\sigma(\mathbf p,\mathbf q) \label{d-p} \\
&=
(2\pi)^{\frac32}\left\lbrace \sqrt{\frac{1}{4\pi}}\langle 1\tfrac12 M \xi |\tfrac12 \sigma \rangle u(q) 
- \sum_{K_3L_3}\langle 1\tfrac12 M \xi|\tfrac32 K_3 \rangle \langle 2
\tfrac32 L_3 K_3 |\tfrac12 \sigma\rangle Y_{2L_3}(\widehat
q)w(q)\right\rbrace\,, \nonumber
\end{eqnarray}
where $\sqrt 3$ is the spectroscopic factor, 
$\psi_{d}(M,\mathbf p)$ is 
the deuteron wave function in 
momentum space, 
$M$ and $\xi$ are spin projections of the deuteron and the proton and
\be
\begin{split}
u(q)&=\sqrt{3}\int_0^\infty dp\, p^2 \left[
u_d(p)\psi_{2}
(p,q) + w_d(p)\psi_{4}
(p,q)
\right],\\
w(q)&=-\sqrt{3}\int_0^\infty dp\, p^2 \left[
u_d(p)\psi_{3}
(p,q)  +
 w_d(p)\psi_{5}
(p,q) \right]\  {;}
\end{split}
\label{WFD}
\ee
here $u_d(p)$ and $w_d(p)$ are the deuteron $S$ and $D$ wave functions,
respectively\,\footnote{ For the convention given by
Eq.~(\ref{alternative_convention}) one must replace $w(q)$ by $-w(q)$. This notation
was used, e.g., in Ref.~\cite{Kobushkin_Deuteron-93}.}. The momentum distribution of
the deuteron in $^3$He is $d(q)=u^2(q) + w^2(q)$.

The effective numbers of the deuterons in $^3\mathrm{He}$, $\mathcal N_d=\int
d^3q\, q^2 d(q)$, are 1.39 and 1.36 for the Paris and CD-Bonn potentials.
These can be compared with $\mathcal N_d=1.38$ obtained in variational
calculations~\cite{Schiavilla} with both the Argonne and Urbana potentials.
The probabilities of the $D$-wave in the $d+p$ configuration are
1.53\,\% and 1.43\,\% for the Paris and CD-Bonn potentials, respectively.

\section{\label{Sec:Spin}Spin-dependent observables}
\subsection{\boldmath Tensor analyzing powers and the $D_2$ parameter}

In a plane wave Born approximation the tensor analyzing powers
$T_{20}$, $T_{21}$ and $T_{22}$ of the $(d,t)$ and $(d,^3$He) reactions at
low energies are determined by a single parameter, $D_2$, used, for example, in
Refs.~\cite{Knutson,BH,Roman,Sen}: $D_2=\lim_{q\to 0}w(q)/[q^2u(q)]$,
i.e., $w(q)/u(q)\approx q^2D_2$ at small $q$.
The $D_2$ parameter is closely related
to the asymptotic $D$ to $S$ ratio for the $p+d$ partition
of the $^3$He wave function, as is noted in Ref.~\cite{Sen}.

The spin-dependent observables considered here depend upon the bilinear forms
of $S$ and $D$ waves of the $^3$He wave function and the behavior of their
ratio at small $q$ is completely governed by the $D_2$ parameter. In
Table~\ref{tab:D2} we compare this parameter, calculated for the bound $3N$ system
(using the wave functions based on different potentials) with the value
derived from experiment.
\vspace{-0.5cm}
\begin{table*}[h] 
\caption{\label{tab:D2}$D_2(3N)$ parameter (in fm$^2$).}
\begin{center}
\begin{tabular}{lllll}
  Paris   &  CD-Bonn  & AV18 \cite{Schiavilla}&  Urbana \cite{Schiavilla}& experiment \cite{Sen}\\
\hline
-0.2387   &  -0.2487  & -0.27                 & -0.23                    & -0.259$\pm$0.014
\end{tabular}
\end{center}
\vspace{-1.cm}
\end{table*}

\subsection{Tensor polarization of the deuteron}
We start by considering the tensor polarization $\rho_{20}$ of the deuteron in
($^3$He,$d$) breakup.
The quantization axis is chosen along the deuteron momentum, i.e.,
$\mathbf q=(0,0,-q)$. We obtain (see also Ref.~\cite{3hekobeas} for details)
within the spectator model, that
\be\label{rho20.2}
\rho_{20}=-\frac{1}{\sqrt{2}}\frac{2\sqrt2 u(q)w(q)+w^2(q)}{u^2(q)+w^2(q)}
\  {;}\  \text{at small $q$}:\  \rho_{20}\approx -2 \frac{w(q)}{u(q)}=-2q^2D_2\  {.}
\ee
Results of calculations are given  in Fig.~\ref{fig:rho20}
(the left panel). 

Note that even in the case of the breakup of
an unpolarized $^3$He, the deuteron spectator emitted at $0^{\,\circ}$ has a
tensor polarization.

\vspace{-3mm}

\subsection{\boldmath Polarization transfer from $^3$He to $d$}
We consider here the case when the quantization axes for the $^3$He and the
deuteron are parallel and both are perpendicular to the deuteron momentum.
In this case the coefficient of the vector-to-vector
polarization transfer from polarized $^3$He to deuteron is (see Ref.~\cite{3hekobeas})
\be\label{Pol.5}
\kappa_d=\frac23\cdot\frac{u^2(q)-w^2(q)-u(q)w(q)/\sqrt2}{u^2(q)+w^2(q)}\  {.}
\ee

We point out that the expression given in Eq.~(\ref{Pol.5}) differs from
Eq.~(5) of Ref.~\cite{sitdeuteron93}  by a factor 2 (erroneously lost in
Ref.~\cite{sitdeuteron93}).

\begin{figure}[h] \vspace{-3mm}
  \includegraphics[width=0.48\textwidth]{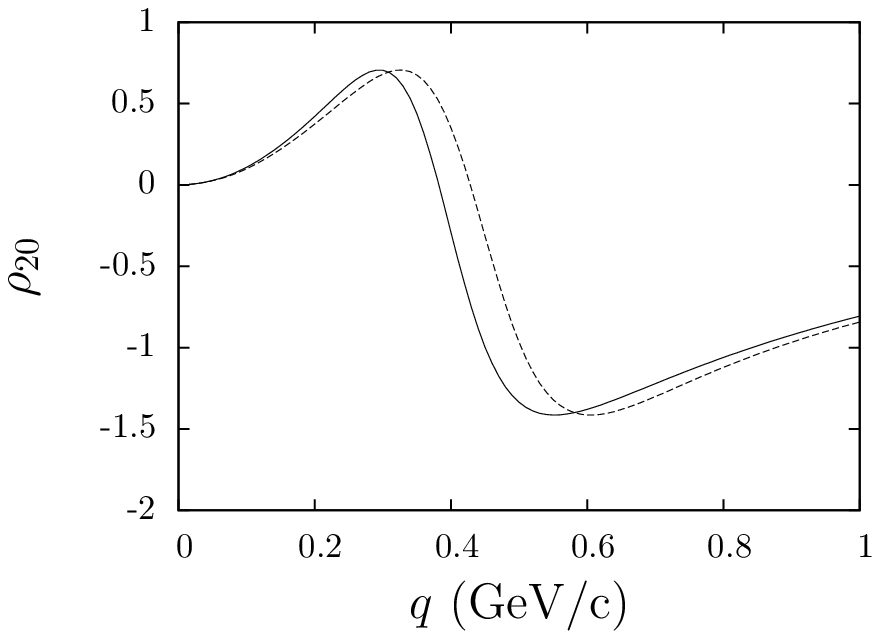}
  \hfill
  \includegraphics[width=0.48\textwidth]{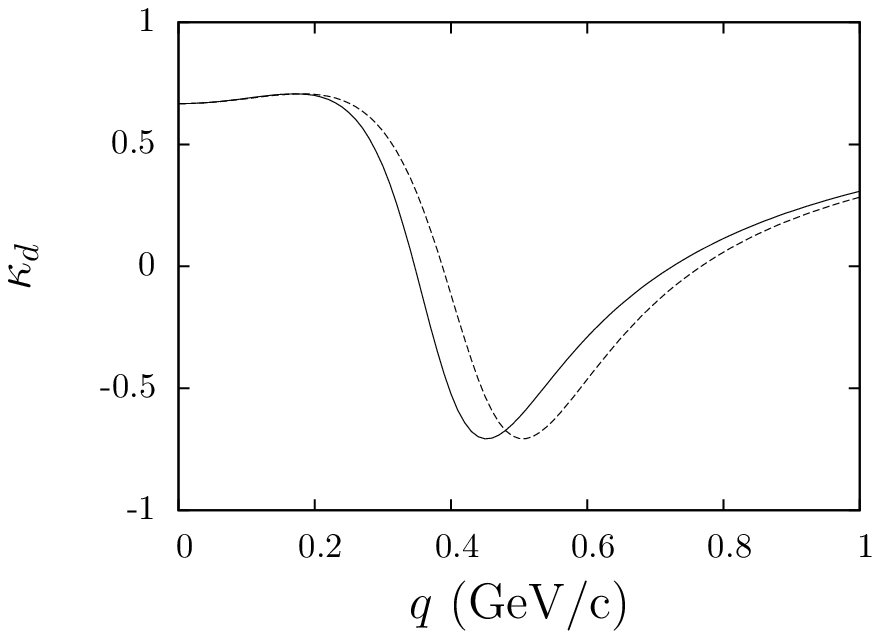}
\caption{\small
Tensor polarization of the deuteron in $^3$He (left) and polarization transfer
$\kappa_d$ from $^3$He to $\mathrm{d}$ (right). Solid and dashed lines are
for the Paris and CD-Bonn potentials.}
\protect\label{fig:rho20}
\vspace{-0.5cm}
\end{figure}

Results of calculations for $\kappa_d$ are shown in Fig.~\ref{fig:rho20}
(the right panel).

The observables $\kappa_d$ and $\rho_{20}$ are related by:
    $\left(\frac{3}{2}\kappa_d\right)^2 + \left(\rho_{20} + \frac{1}{2\sqrt{2}}\right)^2
= 9/8$.
Furthermore, at small $q$
\begin{equation}\label{eq:kappadsmallq}
   \kappa_d\approx
    \frac{2}{3}\left(1-\frac{q^2D_2}{\sqrt{2}}  \right)\approx
\frac{2}{3}\left(1+\frac{\rho_{20}}{2\sqrt{2}}  \right)\  {,}
\  \  \text{i.e.} \ \kappa_d\to 2/3\  \text{when}\  q\to 0\  {.}
\end{equation}

\subsection{\boldmath Polarization transfer from $^3$He to $\boldsymbol{p}$}
The polarization transfer from $^3$He to $p$ is defined by:
\be\label{He-p.1}
\kappa_p=\frac{p_{\frac12\frac12}-p_{\frac12-\frac12}}{p_{\frac12\frac12}+p_{\frac12-\frac12}},
\ee
($p_{\sigma\xi}$ are defined in the subsection~\ref{sec:OneNucl}; details are
in~\cite{3hekobeas}). At $\theta=90^\circ$ this reduces to
\be\label{He-p.2}
\kappa_p=\frac{\rho_1-\rho_2-\rho_4-2(\rho_3+\rho_5)+2\sqrt2(\rho_{13}+\rho_{45})}{\rho_{1}+3(\rho_{2}+\rho_{3}+\rho_{4}+\rho_{5})}\  {,}
\ee
where $\rho_{\mu\nu}(q)=[3/(4\pi)]\int_0^\infty dp\,p^2\psi_\mu(p,q)\psi_\nu(p,q)$.

\begin{figure}[h] \vspace{-3mm}
  \includegraphics[width=0.48\textwidth]{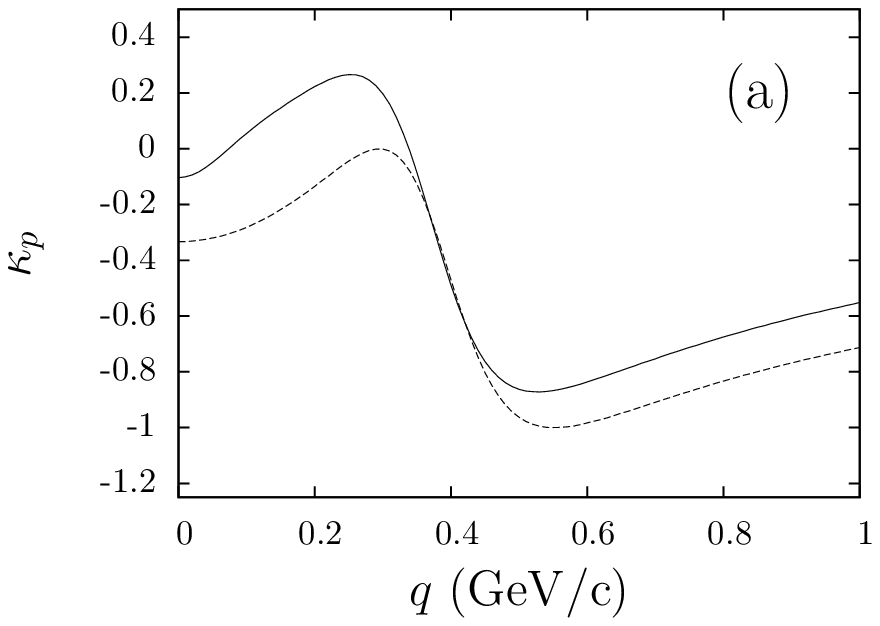} \hfill
  \includegraphics[width=0.48\textwidth]{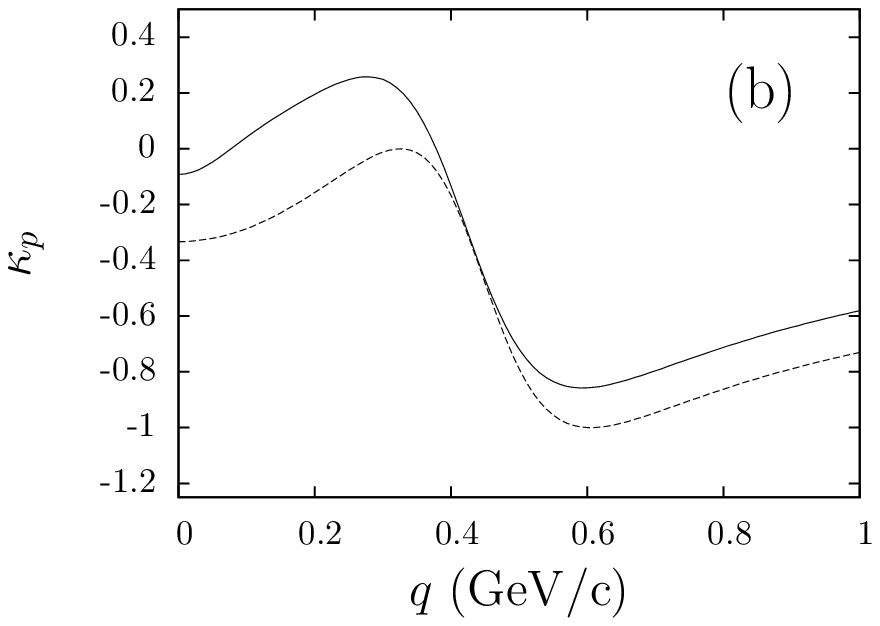}
\caption{\small The coefficient of polarization transfer from $^3$He to the
proton. The $^3$He wave function used is based on the Paris potential (a) and
CD-Bonn potential (b). Solid line: full wave function; short-dashed line:
only the $d+p$ projection (i.e., the $\widetilde \kappa_p$).}
\label{fig:kappastoproton}
\vspace{-0.5cm}
\end{figure}

It is interesting to compare (\ref{He-p.2}) with the polarization transfer for the
$d+p$ projection of the $^3$He wave function (see Fig.~\ref{fig:kappastoproton}):
\be\label{He-p.4}
\widetilde \kappa_p=-\frac{1}{3}\cdot\frac{u^2(q)+2\sqrt{2} u(q)w(q) +2 w^2(q)}{u^2(q)+w^2(q)}.
\ee

It is easy to see that the observables $\widetilde\kappa_p$ and $\rho_{20}$
must be related because they are determined by the ratio of the
two functions $u(q)$ and $w(q)$. One then finds~\cite{3hekobeas}:
\begin{equation}\label{eq:pynrho20}
    \widetilde\kappa_p = -\frac{1}{3}\left(1-\sqrt{2} \rho_{20} \right) {;}\
\text{at small $q$:}\
    \widetilde\kappa_p\approx
    -\frac{1}{3}\left(1- 2\sqrt{2}q^2D_2 \right)\to -\frac{1}{3}\  \text{at}\
    q\rightarrow 0\  {.}
\end{equation}

A linear combination of the two polarization transfer coefficients
at small $q$ is:
\begin{equation}\label{eq:kappasrelat}
1-\left(\widetilde\kappa_p+2\kappa_d\right)\approx
3 q^4 (D_2)^2\approx\frac{3}{4}(\rho_{20})^2\  {.}
\end{equation}

By the way, the similar coefficient of polarization transfer from $^3$He to the
neutron, i.e. $\kappa_n$, is equal to 1 in the spectator model.
\vspace{-3mm}

\section{\label{Sec:Comparing}Comparison with experiment}
\subsection{Empirical momentum distributions}

In order to compare the calculated momentum distributions as well as
the spin-dependent observables with experiment, it
is necessary to establish a correspondence between the argument $\mathbf{q}$ of
the $^3$He wave function and the measured spectator momentum. This must be
done in a way that allows one to take into account relativistic effects in
$^3$He. This problem was discussed in our paper~\cite{3hekobeas}
and here we follow to prescriptions formulated there on the basis of so-called
``light front dynamics''.

Using the corresponding relations, one can extract the relevant momentum
distributions from the measured cross sections; we call such extracted
momentum distributions as ``empirical momentum distributions'' (EMDs) of the
spectators in $^3$He.

\begin{figure}[!h] 
  \includegraphics[height=0.30\textheight]{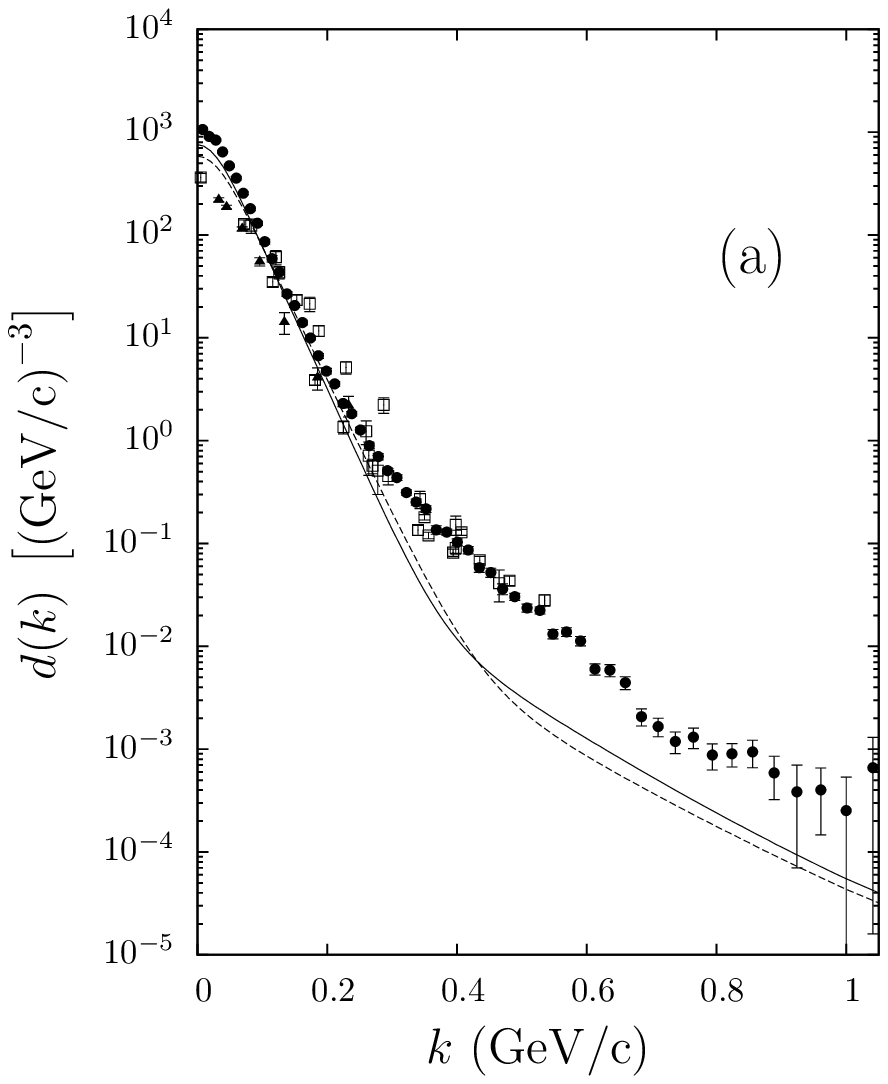} 
\hfill
  \includegraphics[height=0.30\textheight]{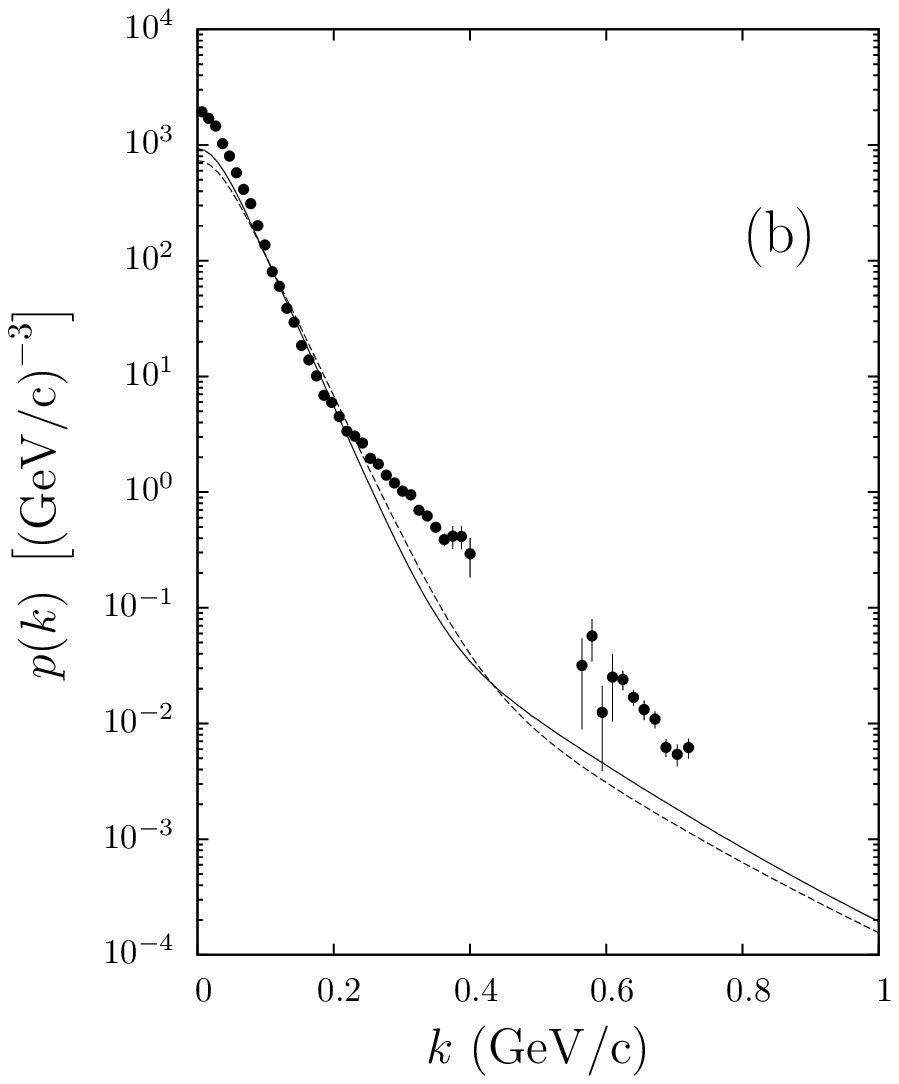}\\
\caption{\small The empirical momentum distributions (EMDs) of deuterons (a)
and protons (b) in $^3$He. The solid and dashed lines are calculated with the
Paris and CD-Bonn potentials. Abscissa: the light cone variable $k$, representing
the argument $q$ of the $^3$He wave function. Full circles: the EMD
extracted from Ref.~\cite{Ableev3He}. Squares and triangles represent data extracted
from Refs.~\cite{Kitching} and \cite{Epstein}. The EMD for protons is normalized to
the calculated one for $k<100$\,MeV/$c$.}
\vspace{-0.5cm}
\protect\label{fig:experiment}
\end{figure}

\begin{sloppypar}
In Fig.~\ref{fig:experiment} we show EMDs for protons and deuterons in $^3$He
extracted from $^{12}$C($^3$He,$p$) and  $^{12}$C($^3$He,$d$) breakup data,
obtained  for fragments, emitted at zero angle and at $p_\mathrm{He}=
10.8$\,GeV/$c$~\cite{Ableev3He}.
They are compared with the results of
our calculations and with available results of other experiments. Good agreement
between the data and the calculations is obvious at
small $k\lesssim 0.25$\,GeV/$c$, which indicates that in this
region the spectator model can be used for data interpretation. Note that the
difference between the light cone variable $k$ and the spectator momentum,
taken in the $^3$He rest frame, is small in this region.
\end{sloppypar}

There is an enhancement of the extracted
EMDs over theoretical curves at very small $k\lesssim 50$\,MeV/$c$.
A natural explanation of this enhancement
appears to be a manifistation of Coulomb effects, which we neglect here, as well as
any possible final state interaction between the outgoing proton and deuteron,
following to~\cite{3hekobeas}.

\begin{sloppypar}
It was argued in Refs.~\cite{Ableev3He} and~\cite{3hekobeas} that
the $k$-variable is an adequate measure for the internal relative momentum of the
$^3$He constituents.
Data on the $(\mathrm{d},p)$ breakup~\cite{JINR(dp)}, including those for
spin-dependent observables~\cite{saclay,dubnacarb} and their analysis, have
resulted in similar conclusions: at small $k\lesssim 0.25$\,GeV/$c$ the
spectator model can be used for the data analysis.
Thus we expect that the reliability of the spectator model 
for the $^3$He breakup at $k\lesssim 250$\,MeV/$c$ should be the same
as in the $(\mathrm{d},p)$ case.
\end{sloppypar}

The data points for momenta above $k\approx 0.25$\,GeV/$c$, where the distances
between the $^3$He constituents become comparable to the nucleon radius or
even less, systematically exceed the calculated momentum distributions.
This is once again very similar to the excess of
data over calculations in the $(\mathrm{d},p)$ breakup~\cite{JINR(dp)}.
It is possible that the observed enhancements in $(^3\mathrm{He},d)$ and
$(^3\mathrm{He},p)$ reactions have the same nature.

\subsection{Tensor polarization of the deuteron}

Data on the tensor polarization $\rho_{20}$ of the deuteron in the reaction
$^{12}\mathrm{C}(^3\mathrm{He},d)$ at several GeV have been
published in~\cite{Sitnik,Sitnikprel}. It should, however, be noted that the
preliminary data~\cite{Sitnikprel} of this experiment have the
opposite sign to those tabulated in the final data set~\cite{Sitnik}.

\begin{figure}[h] \vspace{-3mm} 
  \centering
  \includegraphics[height=0.20\textheight]{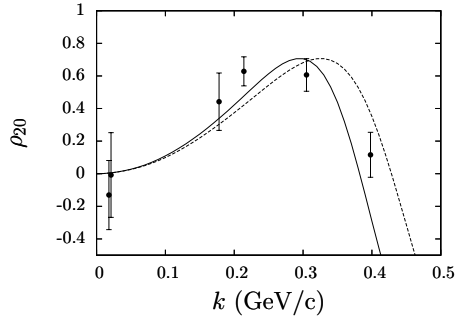}
\caption{\small Deuteron tensor polarization $\rho_{20}$ calculated with
the $^3$He wave
functions for the Paris (solid) and CD-Bonn (dashed) potentials compared with
experimental data. The signs of the data points~\cite{Sitnik} are reversed
to bring them into accordance with the preliminary
results~\cite{Sitnikprel} of the same experiment, as well as with the sign
of experimental data on the $D_2$ parameter for $^3$He. } \label{fig:rho20_dat}
\vspace{-0.5cm}
\end{figure}

On the other hand, the experimental value of the $D_2$ parameter for $^3$He
projected onto the $d+p$ channel has the opposite sign with respect to the
experimental data on the similar $D^d_2$ parameter for the deuteron in the
$n+p$ channel. Therefore the sign of the $\rho_{20}$ under discussion must be
opposite to that of the tensor analyzing power in the $(\mathrm{d},p)$
breakup. Taking this into account, together with the contradiction in signs
of $\rho_{20}$ between Refs.~\cite{Sitnikprel} and \cite{Sitnik}, it is
tempting to conclude that the data tabulated in Ref.~\cite{Sitnik} have the
wrong sign. We therefore use the data from Ref.~\cite{Sitnik} but with a
reversed sign and compare them in Fig.~\ref{fig:rho20_dat} with $\rho_{20}$
calculated according Eq.~(\ref{rho20.2}).

Our results for other spin-dependent observables in the $^3$He breakup cannot
currently be compared with experiment because 
at the present time there are no polarized $^3$He beams with energies
of several GeV/nucleon.

\subsection{Tensor analyzing power in the deuteron breakup}

For the $(\mathrm{d},p)$ breakup reaction with proton emitted at $0^{\circ}$, considered within
the same scheme as in Sect.~\ref{Sec:Spin}, it is possible to connect
corresponding spin-dependent observables with parameter $D^d_2$ defined by the same
equation as for the $^3$He case,
where $u_d(q)$ and $w_d(q)$ functions are the $S$ and $D$ waves of the
bound $p+n$ system. It is straightforward to see that for the analyzing power
$T_{20}$ and the polarization transfer coefficient $\kappa_0$ at small $k$
one has
\begin{equation}\label{eq:T20smallkD2}
     T_{20} \approx -2k^2
     D^d_2\  \  \text{and} \  \
    \kappa_0 \approx
    \left(1+\frac{1}{\sqrt{2}}
    k^2
    D^d_2\right) \approx
 1-\frac{1}{2\sqrt{2}}
 T_{20}\   {.}
\
\end{equation}

\begin{figure}[h] \vspace{-3mm} 
  \includegraphics[origin=br,angle=-90,clip=true,height=0.215\textheight]{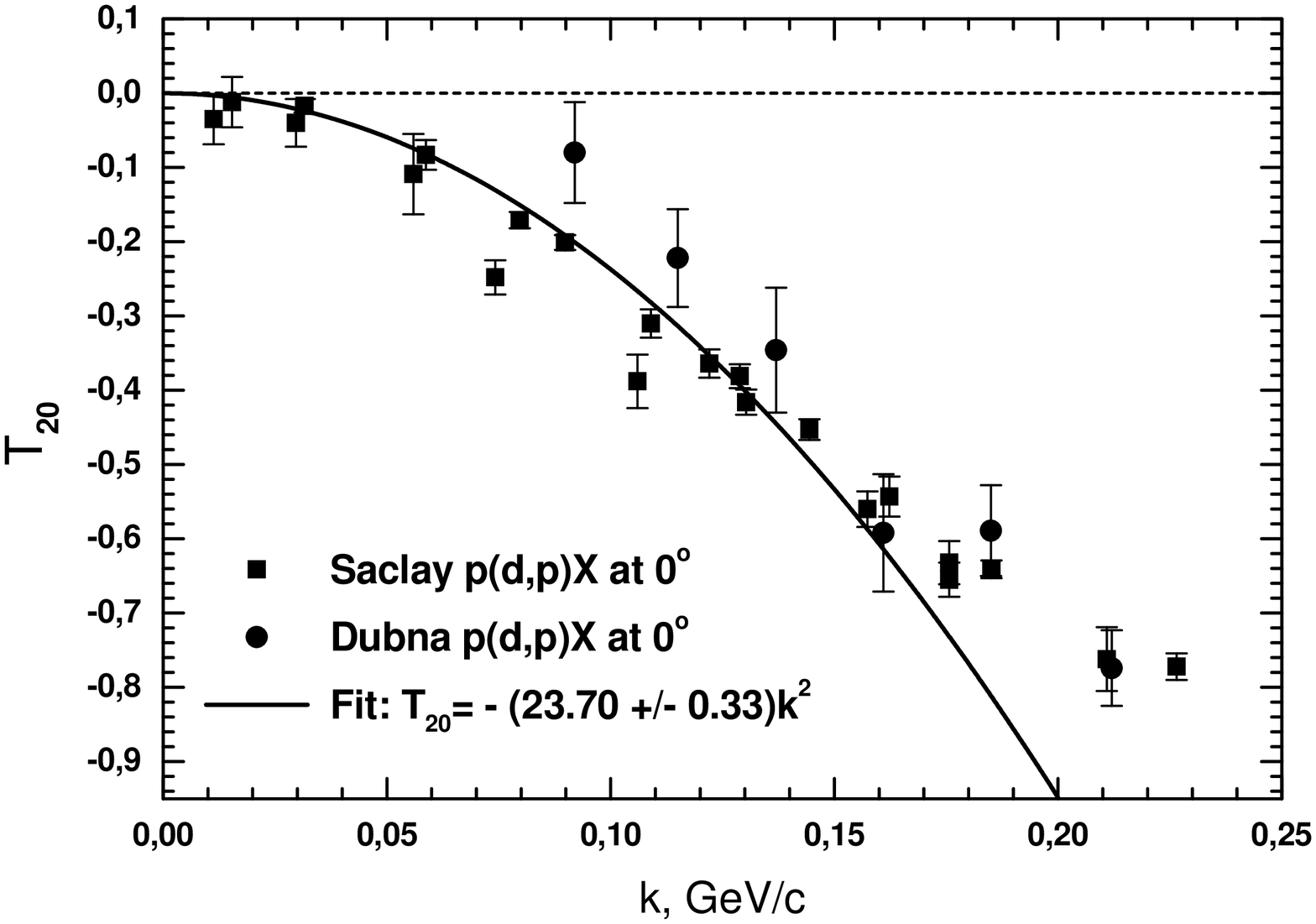}
\hfill
  \includegraphics[origin=br,angle=-90,clip=true,height=0.215\textheight]{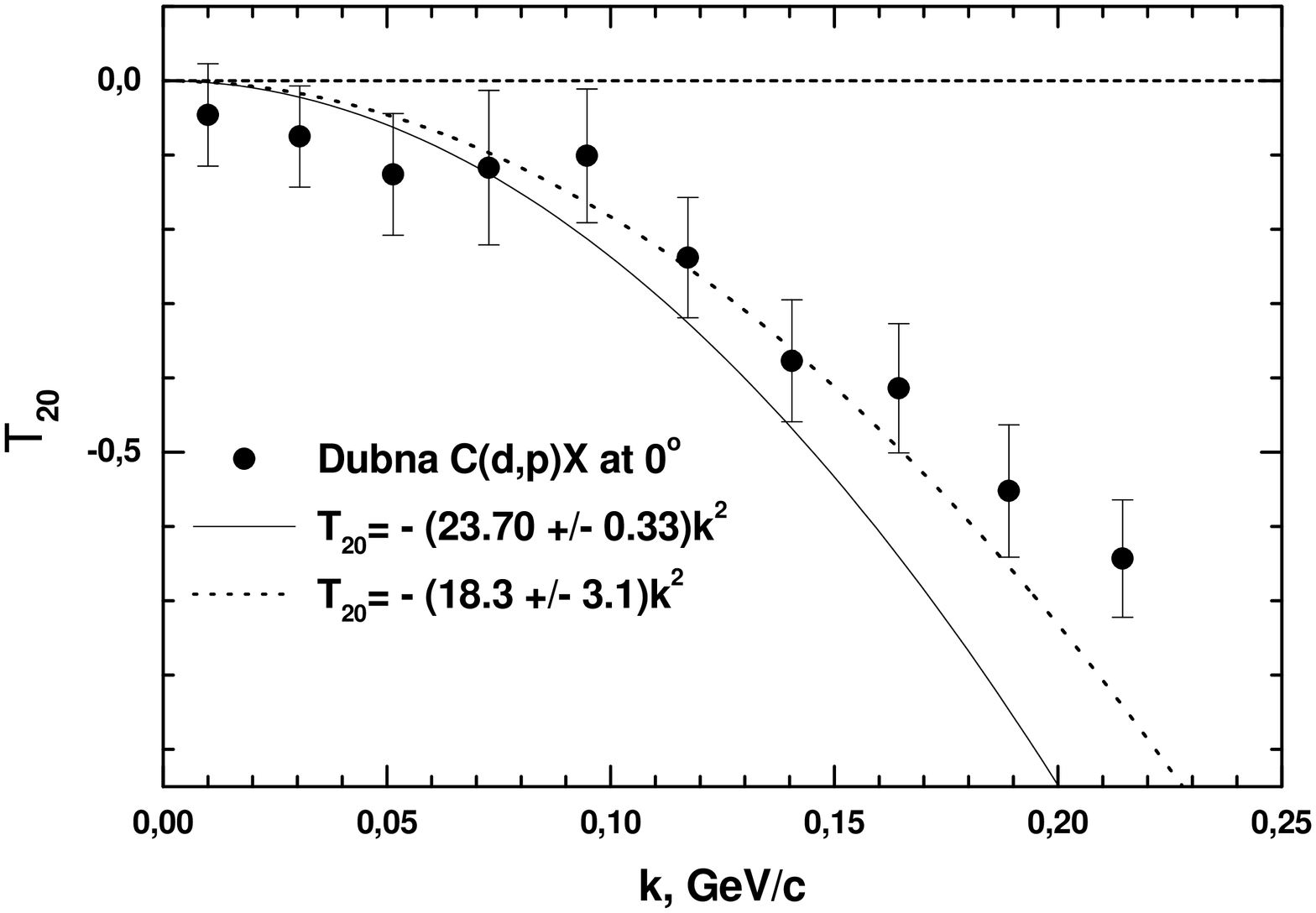}
\parbox[t]{0.48\textwidth}{\caption{\small Data on $T_{20}$ from Refs.~\cite{saclay,dubnacarb} at small $k$.
Solid line: fit according Eq.~(\ref{eq:T20smallkD2}) in the region of
$k\leq 150$\,MeV/c.}\label{fig:t20indeu}}
\hfill
\parbox[t]{0.48\textwidth}{\caption{\small Data on $T_{20}$ from Ref.~\cite{dubnacarb} at small $k$.
The solid line is the same as in Fig.~\ref{fig:t20indeu}
(fixed $D^d_2$). Dotted line: similar fit to the $C(\mathrm{d},p)X$ data
at $k\leq 150$\,MeV/c.}
\label{fig:t20indeucarbon}}
\vspace{-0.5cm}
\end{figure}

The $T_{20}$ data published in~\cite{saclay,dubnacarb} are accurate
enough in order to use Eq.~(\ref{eq:T20smallkD2}) for estimation of the $D^d_2$
parameter (Fig.~\ref{fig:t20indeu}).

\begin{sloppypar}
Fit of the $T_{20}$ data for $p(\mathrm{d},p)X$ reaction in the region of $k\leq 0.15$\,GeV/c
gives $2D^d_2=+(23.70\pm 0.33)$\,(GeV/c)$^{-2}$ with $\chi^2/DoF=19.7/12$
(the Dubna data are not included in the fit as well as two
Saclay data points at $k\approx$\,74 and 106\,MeV/c).
\end{sloppypar}

The obtained value of $2D^d_{2}=+(23.7\pm 0.33)$\,(GeV/c)$^{-2}$ should be compared
with values published in~\cite{Knutson}: $2D^d_2=+(22.19\pm0.82)$\,(GeV/c)$^{-2}$
and in~\cite{deud2}:
$2D^d_2=+(24.80\pm0.67)$\,(GeV/c)$^{-2}$.
Theoretical estimations of this parameter (in (GeV/c)$^{-2}$ units)
can be found, for example, in
papers~\cite{nijmeta},\cite{krasnopol} for different $NN$ potentials (in the paper
by E.~Epelbaum~\cite{nijmeta} the estimations are based on the chiral EFT calculations
in N$^3$LO); all of them are in the interval from $+$24.07 to $+$24.99 with two
exceptions: for RSC potential ($+$25.09 in~\cite{krasnopol}) and the old MSU potential
($+$25.76, see~\cite{krasnopol} as well).

Data for $T_{20}$ in the $C(\mathrm{d},p)X$ breakup from~\cite{dubnacarb} are less
accurate in comparison with the $p(\mathrm{d},p)X$ data from~\cite{saclay},
but still can be used in order to address
the question of the $T_{20}$ sensitivity to Coulomb effects at $k<50$\,MeV/c \cite{kobushemet}. As it is
shown in Fig.~\ref{fig:t20indeucarbon}, these effects (if exist) are invisible at the
present data accuracy. (In both cases we do not take into account any possible systematic
uncertainties of the experiments.)
 \vspace{-3mm}

\section{\label{Sec:Conclusions}Conclusions}

We have presented here an analysis of the spin-dependent observables for
$(^3\mathrm{He},d)$, $(^3\mathrm{He},p)$, and $(\mathrm{d},p)$ breakup reactions
and obtained some rather strict relations between experimental observables
at small internal momenta of fragments.

Our analysis demonstrates that the breakup reactions with the
lightest nuclei at intermediate energies provide a new way for obtaining
experimental data on the $D_2$ parameter for these nuclei, which is
complementary to the usual methods, involving rearrangement reactions at
low energies.

Alternatively, the $(\mathrm{d},p)$ breakup reaction can be used for polarimetric
purposes (for example, measurements of the deuteron beam tensor polarization)
because (i) the accuracy of knowledge of the $D^d_2$ parameter is now high enough
for such purposes and (ii) the cross section of this reaction is high enough,
what results in rather high ``figure of merit'', almost independent upon the
beam energy.

We emphasize that the different conventions regarding the angular momentum
summations for the 3$N$ system result in different forms for the formulae
connecting spin-dependent observables with the $^3$He wave function
components. Of course, the final numerical results do not depend on the
conventions provided that the calculations are  performed systematically
within one chosen scheme. However the occasional mixing of the schemes leads
unavoidably to erroneous results. Therefore an explicit indication of the
chosen angular momentum summation scheme is important for the
applications~\protect\footnote{ Perhaps the lack of such indication
explains, why the sign of the $D$-wave, parametrized in \cite{GW} on the basis
of values tabulated in \cite{Schiavilla}, is opposite to that of the original
tables.}.

Comparing the results of calculations of the deuteron and proton momentum
distributions in the $^3$He nucleus with existing experimental data, we conclude
that the model used for the $^3$He breakup reactions works reasonably well
for $k\lesssim 250$\,MeV/$c$ but at higher momenta the data and calculations
are in systematic disagreement. This disagreement, i.e., the enhancement of
the experimental momentum distributions over the calculated ones above
$k\approx 0.25$\,GeV/$c$ is very similar to the enhancement of data over calculations
observed for the $(\mathrm{d},p)$ fragmentation~\cite{JINR(dp)} at small emission
angles. This was interpreted for the two-nucleon system as a manifestation of
the Pauli principle at the level of constituent quarks~\cite{Kobushkin}. In other
words, an extrapolation to this region of the wave function based on
phenomenological realistic $NN$ potentials for point-like nucleons is
questionable even when relativistic effects are taken into account within the
framework of light cone dynamics. \vspace{-3mm}

\section{Acknowledgments}
{
 The authors gratefully acknowledge Yuri Uzikov and Colin Wilkin for
 their interest to the work and fruitful discussions on various points considered
 in the present paper. The work of  A.P.K. was supported by the Research Programm
``Research in Strong Interacting Matter and Hadron Dynamics in Relativistic Collisions 
 of Hadrons and Nuclei'' of  National Academy of Sciences of Ukraine.} 
 \vspace{-3mm}


\end{document}